\documentclass[twocolumn,prb,aps,epsf,showpacs,superscriptaddress]{revtex4-1}

\usepackage{graphicx}
\usepackage{dcolumn}
\usepackage{bm}

\newcommand{\BrET}{\mbox{$\kappa$-(ET)$_2$Cu[N(CN)$_2$]Br}}
\newcommand{\Br}{\mbox{$\kappa$-(BEDT-TTF)$_2$Cu[N(CN)$_2$]Br}}
\newcommand{\ETX}{\mbox{$\kappa$-(ET)$_2X$}}
\newcommand{\dwave}{$d$-wave}
\newcommand{\tc}{$T_c$}
\newcommand{\I}{\mathrm{i}}
\newcommand{\tbco}{Tl$_2$Ba$_2$CuO$_{6+\delta}$}
\newcommand{\ybco}[1]{YBa$_2$Cu$_3$O$_{#1}$}

\newcommand{\figurewidth}{1}

\begin{document}

\title{In-plane superfluid density and microwave conductivity of the organic superconductor \Br: evidence for {\bf \emph{d}}-wave pairing and resilient quasiparticles}

\author{S.~Milbradt}
\affiliation{Department of Physics, Simon Fraser University, Burnaby, BC, V5A~1S6, Canada}
\author{A.~A.~Bardin}
\affiliation{Centre for Organic Photonics \& Electronics, School of Mathematics and Physics, The University of Queensland,
Brisbane, Queensland 4072, Australia}  
\affiliation{Institute of Problems of Chemical Physics, Russian Academy of Sciences, Chernogolovka, Moscow region 142432, Russia}
\author{C.~J.~S.~Truncik}
\author{W.~A.~Huttema}
\affiliation{Department of Physics, Simon Fraser University, Burnaby, BC, V5A~1S6, Canada}
\author{A.~C.~Jacko}
\affiliation{Institut f\"ur Theoretische Physik, Goethe-Universit\"at Frankfurt, Max-von-Laue-Stra{\ss}e 1, 60438 Frankfurt am Main, Germany}
\author{P. L. Burn}
\author{S.-C. Lo}
\affiliation{Centre for Organic Photonics \& Electronics, School of Chemistry and Molecular Biosciences, The University of Queensland,
Brisbane, Queensland 4072, Australia}  
\author{B.~J.~Powell}
\affiliation{Centre for Organic Photonics \& Electronics, School of Mathematics and Physics, The University of Queensland,
Brisbane, Queensland 4072, Australia}  
\author{D.~M.~Broun}
\affiliation{Department of Physics, Simon Fraser University, Burnaby, BC, V5A~1S6, Canada}

\begin{abstract}
We report the in-plane microwave surface impedance of a high quality single crystal of \Br. In the superconducting state, we find three independent signatures of \dwave\ pairing: i)~a strong, linear temperature dependence of superfluid density; ii)~deep in the superconducting state the quasiparticle scattering rate $\Gamma\sim T^3$; and iii)~no BCS coherence peak is observed in the quasiparticle conductivity. Above $T_c$, the Kadowaki--Woods ratio and the temperature dependence of the in-plane conductivity show that the normal state is a Fermi liquid below $\simeq23$~K, yet resilient quasiparticles dominate the transport up to $\simeq50$~K. 
\end{abstract}
 
 
\maketitle{}

It has been widely argued that the doped Mott insulator describes the essential physics of the cuprates.\cite{Lee:2006de} Similarly, the physics of the \ETX\ salts (ET is an abbreviation of BEDT-TTF) appears to be connected to the bandwidth-controlled Mott transition.\cite{Powell:2011p2232} Thus, it is essential to identify and understand the important similarities and differences between these two classes of quasi-two-dimensional superconductor.   In both the cuprates and the \ETX\ salts, the superconducting critical temperature  is only two orders of magnitude smaller than the Fermi temperature; in this sense, both are high temperature superconductors. 

A broad consensus that the cuprates are \emph{d}-wave superconductors was quickly reached.\cite{Scalapino:1995p741,Annett:1996wf,Orenstein:2000tw,Bonn:2006p572} However, the nature of the pairing state of the \ETX\ salts has taken longer to understand due to the lack of a ``smoking gun" experiment.\cite{Wosnitza:2012fs,Powell:2006ke} Early on there was clear evidence for singlet pairing.\cite{Mayaffre:1995di,Kanoda:1996fb,Miyagawa:2004ce,DeSoto:1995gh}  This, and the low symmetry of the organics, limits the pairing symmetry to be either \emph{s}-wave ($A_{1g}$ representation of the $D_{2h}$ point group) or \emph{d}-wave ($B_{2g}$).\cite{Powell:2006ik}  Early heat capacity experiments suggested \emph{s}-wave pairing,\cite{Elsinger:2000df,Muller:2002hp} but more recent low temperature data points to \emph{d}-wave pairing.\cite{Taylor:2007ca} Measurements of the NMR relaxation rate support unconventional pairing.\cite{Mayaffre:1995di,Kanoda:1996fb,Miyagawa:2004ce,DeSoto:1995gh} Disorder studies show a reduction in  $T_c$ with increasing scattering \cite{Analytis:2006ip,Sasaki:2012vf} but, for larger scattering rates, the suppression of $T_c$ is less than expected for non-\emph{s}-wave superconductors. 
 
 Attempts to locate the nodes expected in a  \emph{d}-wave superconductor  have not yet yielded a simple picture. The in-plane thermal conductivity shows a four-fold angular variation with minima at 45$^\circ$ to the crystal axes;\cite{Izawa:2001gp} whereas when a magnetic field is rotated in the plane both the heat capacity \cite{Malone:2010go} and the millimetre wave absorption \cite{Schrama1999} have minima when the field is aligned with the crystal axes. At first sight these results seem contradictory, but both experiments are extremely difficult to interpret \cite{Vorontsov:2007dn} and a complicated phase diagram could occur as a function of field strength and temperature.\cite{Vorontsov:2007dn,Malone:2010go} Nevertheless, as this has not yet been observed, these measurements have not yet settled the pairing symmetry.  There have also been attempts to directly image the gap via scanning tunneling microscopy (STM).\cite{Ichimura:2008dx} Some care is needed with the interpretation of these experiments as the coherence peaks, the key feature of a superconducting gap, are not observed. Thus a ``V-shaped" differential conductance does not necessarily indicate d-wave superconductivity; similar differential conductances are also found in similar measurements of conventional superconductors with surfaces dirty enough to suppress the coherence peaks.\cite{Bando:1990ij} However, the observation\cite{Ichimura:2008dx} of a zero-bias conductance piece is consistent with the presence of Andreev bound states that might emerge on a rough surface of an unconventional superconductor.	 

  Measurements of London penetration depth and superfluid density, which directly probe the superconducting quasiparticle spectrum,  have further complicated the picture. Several early studies reported data consistent with a nodeless $s$-wave state.\cite{Lang:1992do,Lang:1992co,Yoneyama:2004ji}  On the other hand, a particularly high-resolution study found evidence of low energy excitations,\cite{Carrington:1999p2483} but obtained an anomalous $T^{3/2}$ temperature dependence of the superfluid density. However, the interpretation of these experiments was complicated by their inability to measure the \emph{absolute} penetration depth.  Critically, no measurement to date has reported the linear temperature dependence of in-plane superfluid density expected for a clean $d$-wave superconductor.

In this paper we use microwave surface impedance to probe the in-plane charge dynamics of \BrET, both above and below \tc. (Recall that for \Br~ the $a$-$c$ plane is the highly conducting plane.)  In the superconducting state, the experiment measures the \emph{absolute} London penetration depth, $\lambda_L$, as a function of temperature.  From this we obtain the superfluid density, $\rho_s(T) \equiv 1/\lambda_L^2(T)$.  We observe a strong, linear temperature dependence of $\rho_s$, providing clear evidence of nodes in the energy gap 
consistent with $d$-wave pairing.  The measurements also provide access to the quasiparticle conductivity, $\sigma_1$, from which we extract the quasiparticle scattering rate, $\Gamma$, and the \emph{in-plane} normal-state resistivity, $\rho_\parallel$.  It is important to note that most previous measurements of resistivity in \BrET\ have been measured perpendicular to the highly conducting planes due to difficulties in obtaining properly calibrated in-plane resistivity data.\cite{Dressel:1994vl,Stalcup:1999er,Strack:2005kc}  By providing some of the first reliable measurements of this quantity, the microwave experiment also allows us to perform important tests of dynamical mean field theory (DMFT) \cite{Merino:2000jt,Limelette:2003ca,Deng:2013cs} and of the predicted Kadowaki--Woods ratio in these materials.\cite{Jacko:2009gf}

Single crystals of \BrET \, were grown by controlled electrocrystallization  in a dichloromethane solution containing 8\% (vol.) ethanol.\cite{Anzai:1995jk} Low current densities (0.03--0.21 $\mu$A$\cdot$cm$^{-2}$) and a three-compartment cell were employed to produce high quality single crystals.\cite{Bardin:2012hy} Crystal growth took $\sim5$ weeks and the high quality of the crystals was confirmed by  x-ray crystallography.

\begin{figure}[t]
\centering
\includegraphics[width= \figurewidth\columnwidth]{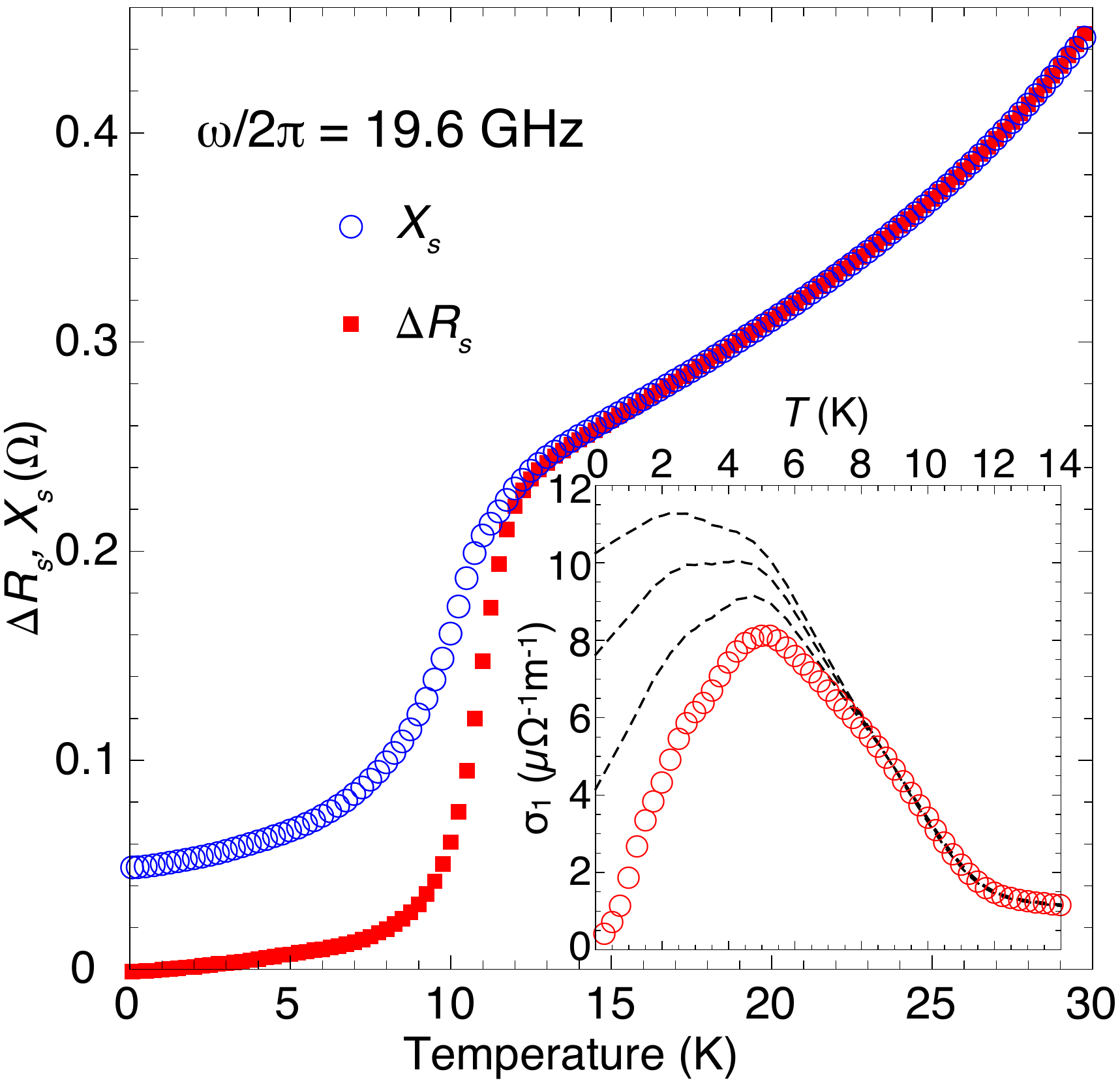}
\caption{(color online).  Temperature dependence of the 19.6~GHz surface impedance, $Z_s = R_s + \I X_s$.  The absolute surface reactance is obtained by finding the temperature-independent offset, $X_{s0}$, that makes $X_s(T) \equiv X_{s0} + \Delta X_s(T)$ match $R_s(T)$ in the Hagen--Rubens regime above $T_c$. Here $X_s(T)$ is plotted on the assumption that residual $R_s$ is zero. Inset: Real part of the microwave conductivity, $\sigma_1(T)$, at 19.6~GHz.  The lowest trace (open circles) indicates  $\sigma_1(T)$ extracted on the assumption that there is no residual surface resistance.  The upper traces (dashed lines) show  $\sigma_1(T)$ for residual $R_s$ of 2, 4 and 6~m$\Omega$, respectively.  The initial rise in $\sigma_1(T)$ on cooling through $T_c$ is robust in the face of uncertainties in residual $R_s$, and signifies a rapid drop in quasiparticle scattering on entering the superconducting state.  Note the absence of a BCS coherence peak below \tc, which would cause $\sigma_1(T)$ to rise almost vertically before falling exponentially at low temperatures.} 
\label{Fig:Zs}
\end{figure}

Surface impedance measurements were carried out at  $\omega/2\pi = 19.6$~GHz using the TE$_{061}$ mode of a rutile dielectric resonator and are plotted in Fig.~\ref{Fig:Zs}.  The measurement system was a dilution-refrigerator-based variant of that described in Ref.~\onlinecite{Huttema:2006p344}, in which the rutile resonator was mounted inside a superconducting enclosure.  A small, platelet single crystal of \BrET, measuring 0.5~mm$\times$0.5~mm$\times $0.1~mm, was attached to one end of a high-purity silicon rod using a small quantity of vacuum grease. During the experiment it was positioned inside the microwave resonator with the microwave $H$ field applied perpendicular to the conducting layers, to induce in-plane screening currents.  The other end of the silicon rod was connected to a temperature-controlled stage \emph{outside} the microwave resonator, allowing sample temperature to be varied in the range 0.075~K to 30~K independently of the resonator temperature, which was kept fixed at 1.6~K during the course of the measurements.  At the beginning of the experiment, the sample was cooled slowly from room temperature at a maximum rate of 1~K/minute. 

Temperature-dependent changes in the sample surface impedance were inferred by cavity perturbation from changes in the resonant frequency, $f_0$, and bandwidth, $f_B$, of the rutile resonator as  the temperature of the sample was varied with respect to base temperature: \mbox{$\Delta R_s +  \I \Delta X_s = \beta (\Delta f_B/2 - \I \Delta f_0)$}.
Here $\beta$ is a resonator constant that depends on the geometry of the sample and the spatial structure of the TE$_{061}$ mode.  In our experiment $\beta$ was determined empirically using a PbSn replica sample of known surface resistance, to an accuracy of better than 5\%.  The determination of absolute $X_s$ is closely related to the determination of absolute penetration depth: $X_s \approx \omega \mu \lambda_L$.  Here we employ a normal-state matching technique that works as follows: above $T_c$,  the normal-state microwave conductivity is predominantly real implying $R_s(T) \approx X_s(T)$;  this condition is imposed and the absolute reactance determined by adding a \emph{temperature-independent} offset to the $\Delta X_s(T)$ data, as shown in Fig.~\ref{Fig:Zs}.  The $R_s(T)$ and $X_s(T)$ data match very well from $T_c$ to 30~K.  This provides an important consistency check, confirming that contributions from thermal expansion and interlayer currents are negligible. Note that in these measurements we have not been able to directly measure the residual surface resistance of the sample, $R_s(0)$: only $\Delta R_s(T)$, its change with temperature.  However, the resulting uncertainty in absolute $X_s$ should be negligible ($R_s(0) \ll X_s(0)$), and the relative error in superfluid density should be small ($\delta \rho_s(0)/\rho_s(0) = \frac{1}{2}R_s(0)/X_s(0) \ll 1$).  The effect of uncertainties in $R_s(0)$ on the microwave conductivity is more substantial, and is shown in the inset of Fig~\ref{Fig:Zs}.

\begin{figure}[t]
\centering
\includegraphics[width= \figurewidth\columnwidth]{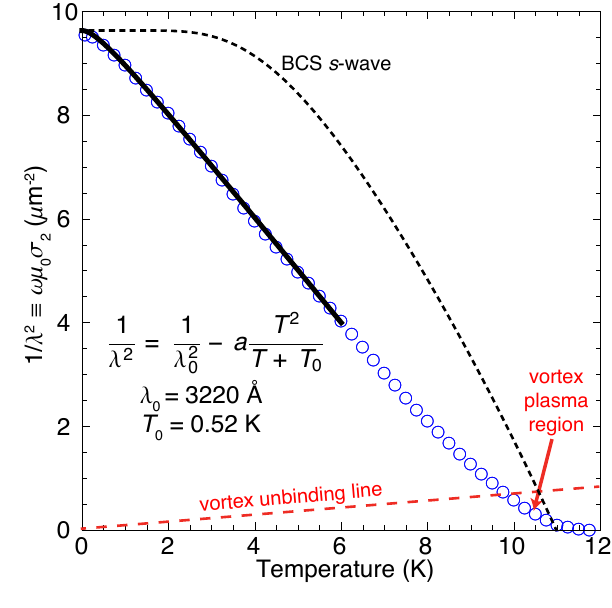}
\caption{(color online).  Superfluid density, $\rho_s(T) \equiv 1/\lambda_L^2(T) = \omega \mu_0 \sigma_2(T)$, obtained from the imaginary part of the microwave conductivity at 19.6~GHz.  $\lambda_0  = 3220$~\AA\ is determined from the \emph{absolute} measurement of $X_s(T)$ in Fig.~\ref{Fig:Zs}, leaving little uncertainty in the shape of $\rho_s(T)$.  The superfluid density displays a strong, linear temperature dependence over most of the temperature range, indicating an order parameter with nodes.  This is strikingly different from the BCS $s$-wave superfluid density \cite{Prozorov:2006wr} (dashed curve).  At very low temperatures $\rho_s(T)$ starts to flatten: the solid line is a fit to a linear-to-quadratic crossover function, $\rho_s(T) = 1/\lambda_0^2 - a T^2/(T + T_0)$, with cross-over temperature $T_0 = 0.52$~K.  The dashed line denotes the expected location of the Kosterlitz--Thouless vortex-unbinding transition.  This should occur when the 2D superfluid density \mbox{$\rho_{\rm s}^{\rm 2D} \equiv  \hbar^2 d/4 k_{\rm B} e^2 \mu_0\lambda_L^2 = (2/\pi)T$}, where $d = 15$~\AA\ is the interlayer spacing.\cite{Geiser:1991jv} $\rho_s(T)$ shows the expected upward curvature close to $T_c$ as it passes through the vortex-plasma regime.} 
\label{Fig:rhos}
\end{figure}

From $Z_s$ we obtain the complex conductivity, $\sigma = \sigma_1 - \I \sigma_2$, using the local electrodynamic expression $\sigma = \I \omega \mu_0/Z_s^2$.  This relation applies when electronic length scales such as in-plane mean free path, $\ell_\|$, and coherence length, $\xi$, are much less than electromagnetic penetration depths, a limit that is satisfied at all but the lowest temperatures in \BrET\ \footnote{The penetration depth is predicted to be affected by nonlocal electrodynamics below a cross-over temperature defined by $k_B T_\mathrm{nl} \approx (\xi_0/\lambda_0) \Delta_0$.\cite{Kosztin:1997p346}  Assuming a zero-temperature in-plane coherence length $\xi_0 \approx 50$~\AA,\cite{Kwok:1990fg,Lang:1994dc,Hagel:1997cq,Wosnitza:2012fs}  $T_\mathrm{nl} \approx 0.35$~K, close to the measured cross-over temperature $T_0$ in $\rho_s(T)$.}. The superfluid density is obtained from the imaginary part of the conductivity using $\rho_s \equiv 1/\lambda_L^2 = \omega \mu_0 \sigma_2$ and is plotted in Fig.~\ref{Fig:rhos}.  We note that our measurement technique, in which we use the normal-state surface impedance as a reference, is able to make an \emph{absolute} determination of $X_s$.  This means that the superfluid density is obtained with very little uncertainty in $\lambda_0$, eliminating  spurious curvature that can be present in $\rho_s(T)$ when only $\Delta \lambda(T)$ is measured.  As seen in Fig.~\ref{Fig:rhos}, $\rho_s(T)$ shows a strong, linear temperature dependence over most of the temperature range, similar to that seen in \tbco\ \cite{Broun:1997p387} and highly underdoped \ybco{6.333}.\cite{Broun:2007p49}  This is the expected behaviour for an order parameter with line nodes in 3D or point nodes in 2D, and is strong evidence for $d$-wave pairing symmetry in \BrET.  At the lowest temperatures there is some slight rounding in $\rho_s(T)$, which we fit using a linear-to-quadratic crossover formula, obtaining a crossover temperature $T_0 = 0.52$~K.  While such behaviour is consistent with $d$-wave superconductivity in the presence of a small density of strong-scattering impurities,\cite{PROHAMMER:1991p557,HIRSCHFELD:1993tf} the inferred value of $T_0$ is also close to where we expect a crossover to nonlocal electrodynamics in the superfluid density,\cite{Kosztin:1997p346} something that is an intrinsic consequence of the nodal structure of a $d$-wave superconductor and must therefore be present in this temperature range. 

From the surface reactance data we obtain a zero temperature penetration depth  $\lambda_0 = 3220$~\AA, corresponding to a superfluid density that is \emph{4 to 40 times larger} than previously reported.\cite{Le:1992ke,Lang:1992co,Dressel:1994vl,Aburto:1998wa,Carrington:1999p2483,Pinteric:2000kd,Pinteric:2002vc,Yoneyama:2004ji}  Since this amounts to a substantial revision, it is critical that we check its validity.  The principle measurement uncertainties in $\lambda$ arise from the surface impedance scale factor and the normal-state matching technique used to determine \emph{absolute} reactance.  In the latter case, we are helped by the fact that the normal-state skin depth at 19.6~GHz is only five times larger than the zero temperature penetration depth, meaning that the relative error in matching is not multiplied by a large factor when transferred to the relative error in $\lambda_0$.  An important test comes from using $\lambda_0$ to estimate the quasiparticle effective mass, $m^\ast = n e^2 \mu_0 \lambda_0^2$.  Taking $n=1.21\times10^{21}$ cm$^{-3}$,\cite{Mielke:1997gb,Geiser:1991jv,Powell:2004ey}  and assuming all electrons condense at low temperature, we obtain $m^\ast = 4.4\pm0.4m_e$.  This can be compared with the thermodynamic mass, $m_\mathrm{th}^\ast = 3 \hbar^2 \gamma/\pi N_A A k_B^2$, where $\gamma$ is the linear coefficient of specific heat (in J/mol.K$^2$), $N_A$ is Avogadro's constant and $A$ is the area per molecular unit.  Experimental values for $\gamma$ lie in the range 22 to 28~mJ/mol$\cdot$K$^{2}$,\cite{Andraka:1991eg,Elsinger:2000df,Taylor:2007ca} giving a combined estimate $m_\mathrm{th}^\ast = 4.5\pm0.25 m_e$.  Separately, quantum oscillation studies have measured the cyclotron mass of the magnetic breakdown orbit in \BrET: combining results from Refs.~\onlinecite{Mielke:1997gb} and \onlinecite{Weiss:1997ck} gives $m_c = 5.44\pm0.1 m_e$.  To the extent that discrepancies with our measurement are significant, we note that Fermi surface anisotropy acts to decrease $m^\ast$ relative to $m_\mathrm{th}^\ast$ and $m_c$.\cite{Merino:2000is}  Another consistency check comes from using our data to predict the expected location of  the Kosterlitz--Thouless--Berezinskii vortex-unbinding transition.\cite{Berezinskii:1972vk,Kosterlitz:1973p177,Ioffe:2002p631,Herbut:2004p735} This is carried out in Fig.~\ref{Fig:rhos}, where we plot the intersection of $\rho_s(T)$ with the vortex-unbinding line, along which the 2D superfluid density \mbox{$\rho_{\rm s}^{\rm 2D} \equiv  \hbar^2 d/4 k_{\rm B} e^2 \mu_0\lambda_L^2 = (2/\pi)T$} in each conducting layer. The relatively high value of superfluid density inferred from our measurements means that vortex unbinding should occur very close to $T_c$.  We observe upward curvature of $\rho_s(T)$ in this vicinity, consistent with the superfluid density becoming frequency dependent as the sample enters the vortex-plasma regime, an effect that is prominent in cuprates.\cite{Corson:1999p716,Bilbro:2011jj}  A similar analysis carried out on previously published data,\cite{Le:1992ke,Lang:1992co,Dressel:1994vl,Aburto:1998wa,Carrington:1999p2483,Pinteric:2000kd,Pinteric:2002vc,Yoneyama:2004ji} in which $\rho_s(T)$ is 4 to 40 times smaller, predicts vortex unbinding in the range 3 to 8~K.  This is not observed.  Finally, our data have recently been shown to be consistent \cite{Dordevic:2013ep} with the Homes scaling law \cite{Homes:2004p354} relating superfluid density to the product of normal-state conductivity and \tc.

\begin{figure}[t]
\centering
\includegraphics[width= \figurewidth\columnwidth]{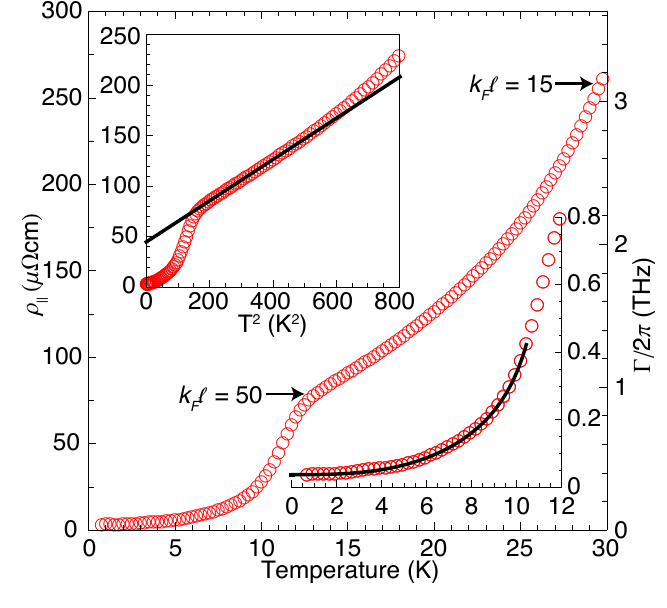}
\caption{(color online).  In-plane resistivity, $\rho_\parallel(T)$, and quasiparticle scattering rate, $\Gamma(T)$.  Above \tc, $\rho_\parallel(T)$ grows initially as $T^2$, steepening with increasing temperature.  \mbox{$k_F \ell_\| \approx 50$} at \tc, falling to 15 by 30~K, where $\ell_\|$ is the in plane mean free path and $k_F$ is the Fermi wave vector. Upper inset: $\rho_\parallel(T)$ vs.\ $T^2$.  Straight line fit is $\rho_\parallel(T) = \rho_0 + A_\parallel T^2$, with $\rho_0 = 43$~$\mu\Omega$cm and \mbox{$A_\parallel = 0.211$~$\mu\Omega$cm/K$^2$}.  Lower inset: $\Gamma(T)$ in the superconducting state, fit to $\Gamma(T) = \Gamma_0 + B T^3/\Delta^2(T)$, with $\Gamma_0/2 \pi = 47$~GHz.  $\Delta(T) = \Delta_0(2.7 \sqrt{T_c/T -1})$ approximates the temperature dependence of the superconducting gap.\cite{Duffy:2001dg}
} 
\label{Fig:Gamma}
\end{figure}

The real part of the microwave conductivity is plotted in the inset of Fig.~\ref{Fig:Zs}.  As shown, the low temperature form of $\sigma_1(T)$ is sensitive to $R_s(T \to 0)$, but the higher temperature behaviour is largely unaffected.  This means that the initial rise in $\sigma_1(T)$ on cooling through $T_c$ is a robust observation and it implies a rapid drop in quasiparticle scattering on entering the superconducting state, which we plot in Fig.~\ref{Fig:Gamma}.  Similar behaviour was originally observed in the cuprate superconductors \cite{Nuss:1991wj,Bonn:1992fx} and has subsequently been seen in materials such as the heavy fermion system CeCoIn$_5$.\cite{Ormeno:2002p404,Truncik:5Z04Ie9v}  The absence of a BCS coherence peak in $\sigma_1(T)$ immediately below \tc\ provides further confirmation of non-$s$-wave pairing.

To examine the scattering dynamics more closely, we extract the quasiparticle scattering rate, $\Gamma(T)$, using a two-fluid model for the complex conductivity, \mbox{$\sigma_1 -\I \sigma_2 = \epsilon_0 \omega_p^2 [f_s/\I \omega + f_n/(\I \omega + \Gamma)]$}, in which the superfluid and normal fractions satisfy the sum rule \mbox{$f_s + f_n = 1$} and $\omega_p = c/\lambda_0$ is the plasma frequency.\cite{Waldram:1997p379}  From this, $\Gamma(T) = \omega [\sigma_2(0) - \sigma_2(T)]/\sigma_1(T)$. Note that the expression for $\Gamma$ is a \emph{ratio} of conductivities, and is therefore insensitive to uncertainties in surface-impedance calibration. The scattering rate data are plotted in Fig.~\ref{Fig:Gamma}.  At \tc, the scattering rate is several times the thermal energy, similar to the situation in optimally doped cuprates \cite{VARMA:1989p468} and CeCoIn$_5$.\cite{Petrovic:2001p391,Ormeno:2002p404,Truncik:5Z04Ie9v}  On cooling through \tc, $\Gamma(T)$ drops rapidly, indicating that the spectrum of fluctuations responsible for inelastic scattering is of electronic origin, in contrast to the phonon fluctuations of a conventional metal.   At lower temperatures, $\Gamma(T) \sim \Gamma_0 + B T^3/\Delta^2$.  This behaviour is characteristic of nodal quasiparticles undergoing large-momentum-transfer scattering processes, either due to the exchange of antiferromagnetic spin fluctuations,\cite{Quinlan:1994vj,Duffy:2001dg} or from direct, short-range repulsion.\cite{Walker:2000ux,Dahm:2005bv}  Interestingly, the energy threshold for exciting Umklapp processes  appears to be small,\cite{Walker:2000ux} suggesting that the gap nodes are separated by approximately a reciprocal lattice vector. 

Because of the low ($D_{2h}$) point group symmetry of \Br\ crystals the above results give us significant insights into the details of the pairing symmetry. For such anisotropic crystals it is generally assumed that the nodes of the order parameter are perpendicular to the highly conducting planes. If this is the case, then there are only two possible irreducible representations to which the order parameter can belong: $A_{1g}$ or $B_{2g}$.\cite{Powell:2006ik} Any nodes in the trivial $A_{1g}$ representation must be accidental and therefore the system can lower its energy by admixing an `$s$-wave' component. We find no signatures of this (in particular there is no coherence peak in the conductivity) suggesting that the order parameter does not belong to the trivial representation.  It must therefore transform as the $B_{2g}$ representation, with nodes along the $a$ and $c$ crystallographic axes. This is often referred to as $d_{xy}$ pairing in the experimental literature, but as $d_{x^2-y^2}$ pairing in the theoretical literature, as the most widely studied anisotropic triangular lattice models have their unit cells rotated with respect to the crystallographic unit cell.\cite{Powell:2006ke}

Above \tc, microwave measurements provide a contactless measurement of the in-plane resistivity, $\rho_\parallel = 1/\sigma_1$.\cite{Dressel:1994vl}   Properly calibrated measurements of $\rho_\parallel$ are difficult to make by conventional means in the \ETX\ compounds, due to their large electrical anisotropy.\cite{Stalcup:1999er,Strack:2005kc}  From the new microwave data, a number of key quantities can now be extracted.  In addition, we can perform a test of DMFT, which provides a powerful framework for understanding the normal state of organic superconductors.\cite{Powell:2006ke,Merino:2000jt,Limelette:2003ca,Deng:2013cs} For these materials, DMFT predicts two well-separated temperature scales. $T_\mathrm{MIR}$ is defined by the Mott-Ioffe-Regel limit, i.e., where $\ell_\|$ equals a lattice constant or $k_F\ell_\|=1$. Above $T_\mathrm{MIR}$ \Br\ is a `bad metal', characterised by the absence of quasiparticles. Below a second temperature scale,  $T_\mathrm{FL}<T_\mathrm{MIR}$, the electrons form a Fermi liquid. In the intermediate regime, $T_\mathrm{FL}<T<T_\mathrm{MIR}$, DMFT predicts that, although there is not a true Fermi liquid, the quasiparticles are resilient and continue to dominate the transport.\cite{Deng:2013cs}

To test these predictions, $\rho_\parallel(T)$ is plotted vs.\ $T^2$ in the upper inset of Fig.~\ref{Fig:Gamma}, showing that quadratic temperature dependence, a key signature of a Fermi liquid, is indeed observed.  The straight-line fit shows $\rho_\parallel(T) = \rho_0 + A_\parallel T^2$, with $\rho_0 = 43$~$\mu\Omega$cm and $A_\parallel = 0.211$~$\mu\Omega$cm/K$^2$.  The magnitude of $\rho_\parallel(T)$ further confirms that the low temperature metallic state is indeed a Fermi liquid. The value of the $A_\parallel$ coefficient can be compared with a recent prediction of the Kadowaki--Woods ratio in strongly correlated 2D local Fermi liquids (the state predicted by DMFT for $T<T_\mathrm{FL}\simeq23$~K):\cite{Jacko:2009gf}
\vspace{-3mm}
\begin{equation}
\vspace{-3mm}
\frac{A_\parallel}{\gamma^2} = \frac{81 \hbar}{4 k_B^2 e^2} \frac{d}{n^2}\;. \label{eq:kwr}
\end{equation}
Here  $n=1.21\times10^{21}$ cm$^{-3}$ is the electron density\cite{Mielke:1997gb,Geiser:1991jv,Powell:2004ey}   and $d=15$~\AA\  is the interlayer spacing.\cite{Geiser:1991jv} It has been confirmed that this formula is accurate to within a factor of 2 over a wide range of strongly correlated layered metals.\cite{Jacko:2009gf}  For $\gamma=22-28$~mJ/mol$\cdot$K$^{2}$ \cite{Andraka:1991eg,Elsinger:2000df,Taylor:2007ca} Equation~(\ref{eq:kwr}) predicts that $A=0.09 - 0.14~\mu \Omega\cdot${cm/K}$^2$, which agrees with our measured value to within the known accuracy of the theory. Previous contactful measurements found that $A_\|\simeq20 \mu \Omega\cdot$cm/K$^2$;\cite{Yasin:2011cw} such a large value suggests that these measurements may be contaminated by the much larger interlayer resistance. Thus, we conclude that our results represent an accurately calibrated measurement of the in-plane resistivity of \Br. Using the standard expression of resistivity in a 2D metal, $\rho_\parallel = h d/e^2 k_F\ell_\|$, we find $k_F \ell_\| \approx 50$ immediately above \tc, decreasing to 15 by 30 K.   Extrapolating to higher $T$, we estimate that $k_F \ell_\| = 1$ for $T\approx 60$~K, in accord with the DMFT prediction \cite{Merino:2000jt,Limelette:2003ca,Deng:2013cs} and previous experimental estimates.\cite{Powell:2006ke}

Overall, our measurements give a detailed picture of behaviour of electrons in \BrET. Measurements of the quasiparticle scattering rate allow for key tests of DMFT and the unified theory of the Kadowaki-Woods ratio in this material. The consistency with these models confirms that the strong electronic correlations  central to the normal state of these materials are accurately described by these theories. In the superconducting state we have provided three clear pieces of evidence that \emph{d}-wave pairing is realised: the linear temperature dependence of the superfluid density; the absence of a BCS conductivity coherence peak; and the $T^3$ dependence of the quasiparticle scattering rate. The microwave measurements therefore remove the ambiguities arising from previous measurements of the penetration depth and represent an important contribution to the emerging consensus of $d$-wave pairing symmetry in \BrET.

\emph{Note added}: Recently, we learned of related results reported independently.\cite{Perunov:2012wo}

The authors thank A.~Carrington, M. Dressel,  R.~H.~McKenzie, N.~C.~Murphy and S.~S.~Yasin for useful discussions/correspondence.  Research support  was provided by the Natural Science and Engineering Research Council of Canada, the Canadian Foundation for Innovation and the Australian Research Council (projects DP1093224 and DP0878523).\\\\


%

\end{document}